\documentclass[preprint,12pt,authoryear,round,numbers]{elsarticle}

\usepackage{amssymb}
\usepackage{amsmath}
\usepackage{natbib}  
\usepackage{hyperref}         
\usepackage{soul}
\usepackage{booktabs}
\usepackage{url}
\usepackage{longtable}
\usepackage{array}        
\usepackage{multirow}     
\usepackage{ragged2e}     
\usepackage{float}
\usepackage{soul}
\usepackage{titlesec}  
\usepackage{xcolor} 
\journal{Chaos, Solutions & Fractals}

\begin{document}

\begin{frontmatter}

\title{CHAOS AND STABILITY IN THE MARINE TROPHIC NETWORK: THE IMPORTANCE OF INTERACTIONS OVER COMPLEXITY}

\author[inst1,inst2]{Ilaria Cunico\corref{cor1}}

\author[inst1,inst2,inst3]{Guido Occhipinti}

\author[inst4]{Gregor Fussmann}

\author[inst1,inst2]{Paolo Lazzari}

\cortext[cor1]{Corresponding author. email: \texttt{icunico@ogs.it}}

\address[inst1]{OGS-National Institute of Oceanography and Applied Geophysics, Trieste, Italy}
\address[inst2]{NBFC-National Biodiversity Future Center, Palermo, Italy}

\address[inst3]{MARBEC,	Univ Montpellier, CNRS, Ifremer, IRD, Sète, France}

\address[inst4]{Department of Biology, McGill University, 1205 Dr Penfield Avenue, Montréal, QC H3A 1B1, Canada}





\author{} 


\begin{abstract}
Understanding the dynamics of real-world complex networks is crucial for assessing their predictability, resilience, and improving ecosystem management, especially in the context of climate change. The relationship between stability and complexity in ecological networks is still debated in the literature.\\
In this modeling study, we investigate whether a complex marine trophic network — characterized by multiple trophic interactions and environmental constraints — exhibits predominantly stable, periodic or chaotic dynamics. We incorporate the microbial loop into a trophic network model, which includes one to three primary producers, one or two consumers, and up to three trophic levels of predators.  The microbial loop is a key process in which bacteria recycle detritus from higher trophic levels into nutrients available for the growth of primary producers, ensuring mass conservation within the system.  We perform numerical simulations to investigate the network’s dynamic behavior, exploring several configurations by turning off predator-prey links between species and varying the high-dimensional parameter space.\\
Our results show that (i) longer trophic chains and (ii) a higher number of consumers increase system chaoticity, whereas (iii) omnivorous interactions promote stability. Notably, many of the configurations exhibit high percentages of chaotic behavior. Feedback loop analysis suggests that the balance between negative and positive interactions plays a key role in the system’s convergence toward a steady state. \\
This study shows that interactions and feedback, rather than complexity, are key drivers of stability, pointing to the absence of a clear stability–complexity relationship and instead highlighting a stability–interaction dependence. Chaotic dynamics may also play an important role, with potential implications for predictability and ecosystem management.
\end{abstract}




\end{frontmatter}



\section{Introduction}
\noindent Since the early days of theoretical ecology, it has been recognized that very simple nonlinear models are capable of producing highly complex behaviors. This led to the idea that examining such models might be sufficient to investigate the complex environmental variability observed in nature. More recent studies, however, have shown that more complex models can promote stability \citep{berryman1989ecological, fussmann2002food, occhipinti2023marine}, and that this property depends on the detailed structure of the network. These studies therefore question the explanatory power of simple models. On the other hand, ecological systems composed of a very large number of individuals, characterized by random dense interactions and approaching the thermodynamic limit, frequently exhibit complex dynamics, including chaotic behavior \citep{ crutchfield1986chaos,roy2019numerical, rogers2022chaos}.
Moreover, it has been formally shown that the ordinary differential equations framework commonly used to model competing species, if appropriately constructed and when the number of species becomes very large, has sufficient freedom to exhibit arbitrarily rich dynamical behavior—including structurally stable dynamics, periodic solutions, and chaotic behavior \citep{smale1976differential}. \\
Here, we contribute to the debate by examining the stability of a mathematical model that is, in structure and parameterization, consistent with a complex marine trophic network, including the microbial loop and mass conservation.
We focus on a trophic structure typical of lower trophic levels and predators grazing upon it, as used in current medium-complexity models to reconstruct marine primary production at both global and regional scales.
These models are used in operational oceanography to provide forecasts and analysis as in the Copernicus Marine Service (CMS, https://marine.copernicus.eu) or for climate projection as produced by the Coupled Model Intercomparison Project Phase 6 (CMIP6, https://pcmdi.llnl.gov/CMIP6/). \\
In particular, in this work we extend an eight-species aquatic trophic network originally analyzed by \cite{fussmann2002food}. We enforce mass conservation in the system by incorporating the microbial loop, a fundamental process at the base of the food web. The microbial loop plays a critical role in aquatic ecosystems by
recycling detritus from higher trophic levels into available nutrients, thereby supporting primary producers and closing the circle of matter
\citep{hemmings2004parameterizing, azam2007microbial}.\\
Starting with the maximally species-rich and connected network, we systematically deactivate the predator-prey links between different species, exploring all possible network configurations and levels of connectivity. We then analyze the stability-complexity relationship by studying the dynamic behavior of the configurations as a function of the mortality parameters. \\
The results show that stability decreases with longer food chains and a higher number of consumers, but increases with omnivorous links. Moreover, compared to the precursor model \citep{fussmann2002food}, configurations exhibit high instability, showing intrinsic oscillations and chaos. Additionally, the analysis of feedback loops indicates that steady states are reached only when stabilizing negative interactions dominate over destabilizing positive ones in the network.\\
The present study aims to understand the dynamic behavior of a complex marine trophic network and is structured as follows. Section \ref{sec2_methodology} describes the numerical model, simulation set up, and the metrics used to identify chaos. In Section \ref{sec 3: Results}, we show the main results, comparing the dynamic behavior across all network configurations. In Section \ref{sec 4: Discussion}, we discuss the results in light of the existing literature, and we summarize the novelty of this work.

\section{Methodology}
\begin{figure}[H]
    \centering
    \includegraphics[width=1.05\linewidth]{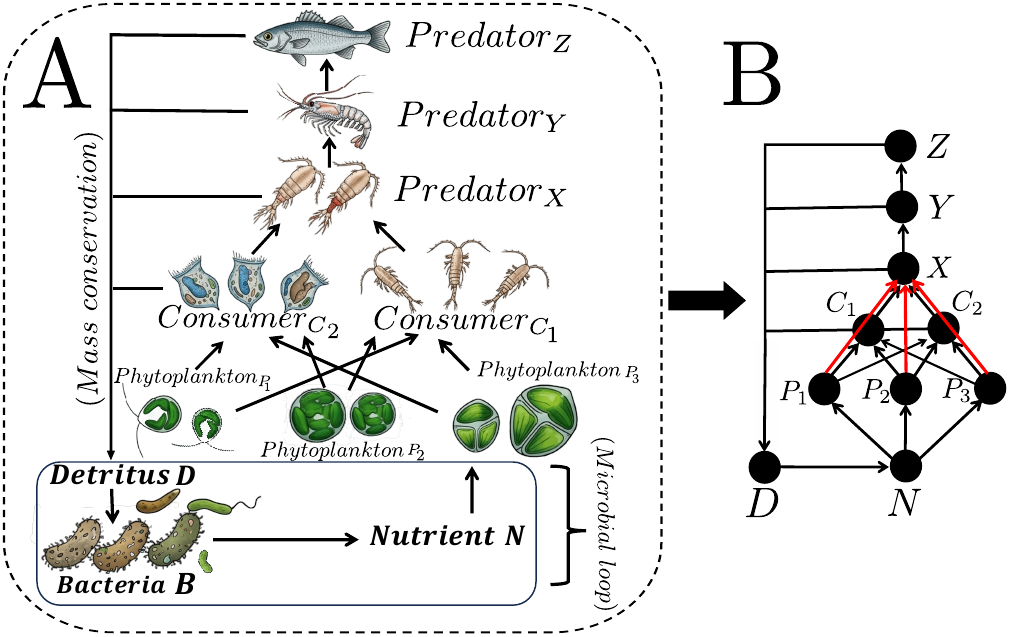}
    \caption{Trophic network structure. The maximally complex network consists of an eight-species marine trophic web, comprising three phytoplankton species $P_1$, $P_2$, $P_3$, two consumers $C_1$, $C_2$, as well as a primary, secondary, and tertiary predator $X, Y, Z$. Additionally, the model incorporates the microbial loop, a fundamental process at the base of the food web, where bacteria $B$ convert detritus $D$ into nutrients $N$, thereby enabling mass conservation. ($A$) Examples of species occupying various trophic levels and ($B$) nomenclature and ecosystem topology. $Red \ arrows$ indicate omnivorous links. Bacteria are not represented, as they are not explicitly modeled.}
    \label{fig:Fig1_model}
\end{figure}
\label{sec2_methodology}
\subsection{Numerical model}
\noindent We extended a trophic network model which includes eight species interacting across five trophic levels \citep{fussmann2002food}, by incorporating real-world environmental constraints, as illustrated in Figure \ref{fig:Fig1_model}. We assume a homogeneous environment without explicit spatial structure, which is more representative of pelagic environments than of highly turbulent coastal waters. Specifically, we integrated the microbial loop, where bacteria recycle detritus from higher trophic levels into nutrients. This process not only supports primary production but also maintains mass conservation within the system \citep{hemmings2004parameterizing}. 
The model is described by the following ordinary differential equations:

\begin{align}
\frac{dP_i}{dt} &= P_i \Big(
\frac{r_i N}{N + K_i} 
- \frac{a_{P_i}^{C_1} C_1}{1 + \sum_{j=1}^3 b_{P_j}^{C_1} P_j} 
- \frac{a_{P_i}^{C_2} C_2}{1 + \sum_{j=1}^3 b_{P_j}^{C_2} P_j} \notag \\
&\quad\quad\quad - \frac{a_{P_i}^{X} X}{1 + \sum_{j=1}^3 b_{P_j}^{X} P_j + b_{C_1}^{X} C_1 + b_{C_2}^{X} C_2}
\Bigg), \quad i=1,2,3 \\
\frac{dC_j}{dt} &= C_j \left(
\frac{\sum_{i=1}^3 a_{P_i}^{C_j} P_i}{1 + \sum_{i=1}^3 b_{P_i}^{C_j} P_i}
- \frac{a_{C_j}^{X} X}{1 + \sum_{i=1}^3 b_{P_i}^{X} P_i + \sum_{k=1}^2 b_{C_k}^{X} C_k} - d_{C_j} \right), \quad j=1,2 \\
\frac{dX}{dt} &= X \left(
\frac{\sum_{i=1}^3 a_{P_i}^{X} P_i + \sum_{j=1}^2 a_{C_j}^{X} C_j}{1 + \sum_{i=1}^3 b_{P_i}^{X} P_i + \sum_{j=1}^2 b_{C_j}^{X} C_j}
- \frac{a_{X}^{Y} Y}{1 + b_{X}^{Y} X}
- d_X
\right) \\
\frac{dY}{dt} &= Y \left(
\frac{a_{X}^{Y} X}{1 + b_{X}^{Y} X}
- \frac{a_{Y}^{Z} Z}{1 + b_{Y}^{Z} Y}
- d_Y
\right) \\
\frac{dZ}{dt} &= Z \left(
\frac{a_{Y}^{Z} Y}{1 + b_{Y}^{Z} Y}
- d_Z
\right) \\
\frac{dD}{dt} &= \sum_{j=1}^2 d_{C_j} C_j + d_X X + d_Y Y + d_Z Z - \frac{1}{\tau} D \\
\frac{dN}{dt} &= \frac{1}{\tau} D - N \sum_{i=1}^3 \frac{r_i P_i}{N + K_i}
\label{eqn:model}
\end{align}

\noindent where $P_i$ represents the primary producers ($i=1,2,3$), $C_j$ the consumers ($j=1,2$), and $X$, $Y$, and $Z$ represent the predators. Omnivorous links involving phytoplankton were restricted to predator X. Links from predator Z were excluded because consumers at that trophic level are generally too large to feed directly on phytoplankton \citep{cohen1993body}, while Y–phytoplankton interactions were omitted to reduce the number of network configurations. Primary producers grow following a Monod growth equation, dependent on nutrients $N$, growth rates $r_i$, and saturation constants $K_i$. Consumers and predators follow a type II Holling equation, where $a_{P_i}^{C_j}$, $a_{P_i}^{X}$, $a_{C_j}^{X}$, $a_{X}^{Y}$, $a_{Y}^{Z}$ are predation rates and $b_{P_i}^{C_j}$, $b_{P_i}^{X}$, $b_{X}^{Y}$, $b_{C_j}^{X}$, $b_{Y}^{Z}$ are the saturation parameters. In addition to predation, consumers and predators have intrinsic mortality rates, denoted by $d_{C_j}$, $d_X$, $d_Y$, and $d_Z$, respectively. 
$D$ represents detritus originating from higher trophic levels, which is remineralized by bacteria (not explicitly modeled) into nutrients on a characteristic remineralization timescale $\tau$ \citep{hemmings2004parameterizing}. The model is implemented using the framework for aquatic biogeochemical modelling \citep[FABM; ][]{bruggeman2014general} and the code can be found in the GitHub repository (\url{https://github.com/icunico/Fussmann_ml.git}). The parameter values and initial conditions have been chosen with biological meaning (i.e., they are consistent with real marine ecosystems, but are not empirical values representative of concrete species) and are provided in Table \ref{tab:parameters} of the \ref{Appendix: Model parameters}. The predation rates and saturation parameters are chosen according to \citeauthor{levin1970community}'s (\citeyear{levin1970community}) rule, since the presence of a single nutrient limits the coexistence area of species.

\label{sec: Methodology}

\subsection{Analysis and model setup}
\noindent We investigated the dynamic behavior of several configurations within the eight-species trophic network introduced in the previous section, systematically removing predator-prey links between species to compare their dynamic behavior.
The network configurations are classified as shown in Fig \ref{fig:network_configurations} of the \ref{Appendix: Nomenclature}. For each configuration, we assess the system's behavior and categorize the outcomes as follows: (i) extinction, where at least one species drops below the survival threshold; (ii) steady state, where all species reach a steady state; (iii) periodicity, where at least one species exhibits periodic oscillations; or (iv) chaotic behavior, where at least one species shows chaos. 
To compare the dynamic behavior across different configurations, all parameters are fixed, except for the mortality rates. Due to the 'curse of high dimensionality' \citep{altman2018curse}, exploring all parameters of the system is not feasible because of the high computational cost associated with evaluating all parameter combinations. Nevertheless, the self-regulation feedback of the mortality rate is a crucial parameter for investigating the stability of trophic networks \citep{barabas2017self, barbier2019pyramids}. Additionally, our model does not incorporate external perturbations or stochastic factors, as the primary aim is to determine whether the complexity emerges from the intrinsic dynamics of the network. \\
 
\subsection{Stability, Periodicity, and Chaos Metrics}
\noindent To define the network's system behavior, we performed simulations representing a period of 20 years, discarding the first 2 years to eliminate the transient phase. We determined whether the system reached a steady state by evaluating the ratio of the standard deviation to the mean for all species. The system was considered steady if the maximum ratio was below $ \delta=0.01$ for each species, and oscillatory if at least one species exceeded this survival threshold, set to 0.001. \\
To distinguish between chaotic and periodic oscillations, we applied two information theory quantifiers: the normalized Shannon entropy and the statistical complexity measure \citep{zunino2012distinguishing, rosso2007distinguishing}. These metrics characterize, respectively, the diversity and the correlational structure of a time series. The dynamical behavior of the system was further examined by analyzing the multiscale complexity-entropy causality plane. Chaos is associated with intermediate values of normalized Shannon entropy and the maximum complexity measure \citep{zunino2012distinguishing}.  These indices were computed using the Python package ordpy \citep{pessa2021ordpy}. To assess the robustness of these methods, we replicated the results of \cite{fussmann2002food}, who applied the traditional maximum Lyapunov exponent. Our findings are consistent with theirs; further details can be found in \ref{Appendix: Us vs Fussmann}, Figure \ref{fig:comparisionFussmann}. \\
For each configuration, we simulated 100.000 parameter combinations by modifying the mortality rates using the PARSAC tool \citep{bolding2020parsac}. The PARSAC tool runs simulations in parallel, reducing the computational cost.

\section{Results}
\label{sec 3: Results}
\noindent We compared the dynamic behavior across network configurations that differed in topology, species number, and connectivity. Detailed information is provided in Table \ref{table:Results} of the \ref{Appendix: Simulation results}. 
\begin{figure}[H]
    \hspace{-50pt}
\includegraphics[width=1.2\linewidth]{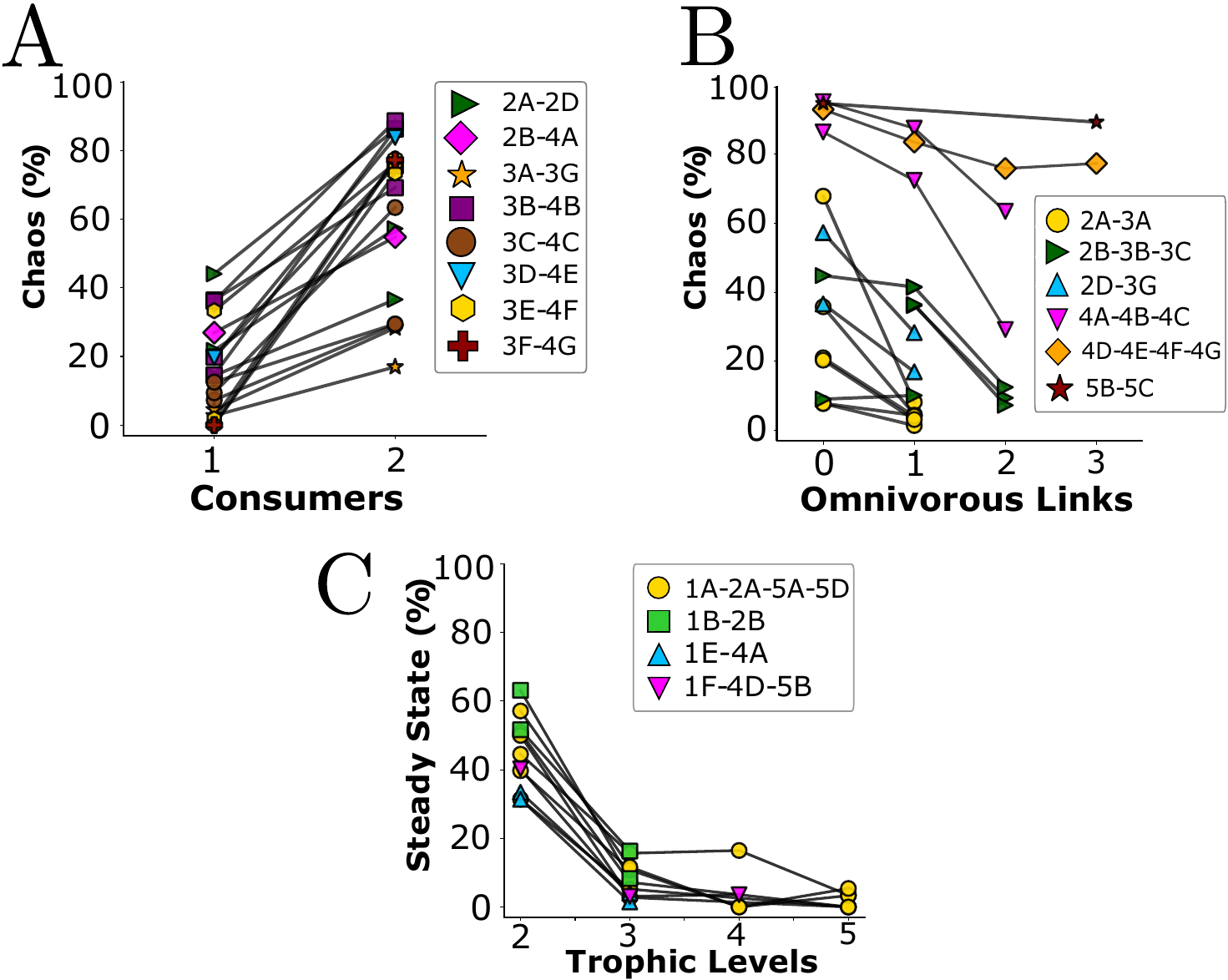}
    \caption{Comparison of the dynamic behavior across significant network configurations. Each point represents a specific configuration, with identical colors representing networks that share the same topology. Legend nomenclature corresponds to that depicted in the \ref{Appendix: Nomenclature} (\ref{fig:network_configurations}). $(A)$ Chaos percentage as a function of the number of consumers. $(B)$ Chaos percentage as a function of omnivorous links. $(C)$ Steady state percentage as a function of trophic levels.}
    \label{Summary Results}
\end{figure}
\noindent In brief,  Figure \ref{Summary Results}A shows the percentage of chaos as a function of the number of consumers, Figure \ref{Summary Results}B shows the percentage of chaos as a function of the number of omnivorous links, and Figure \ref{Summary Results}C depicts the percentage of steady states as a function of the number of trophic levels. Configurations with very small coexistence regions are omitted, as they are not representative. In all panels, each point represents a configuration, and identical colors indicate networks that share the same topology, with the nomenclature corresponding to that described in Figure \ref{fig:network_configurations} of the \ref{Appendix: Nomenclature}. Since multiple realizations can exist for the same configuration, multiple points of the same color are shown.

\begin{figure}[H]
    \vspace{-35mm}
    \centering
    \includegraphics[width=1.0\linewidth]{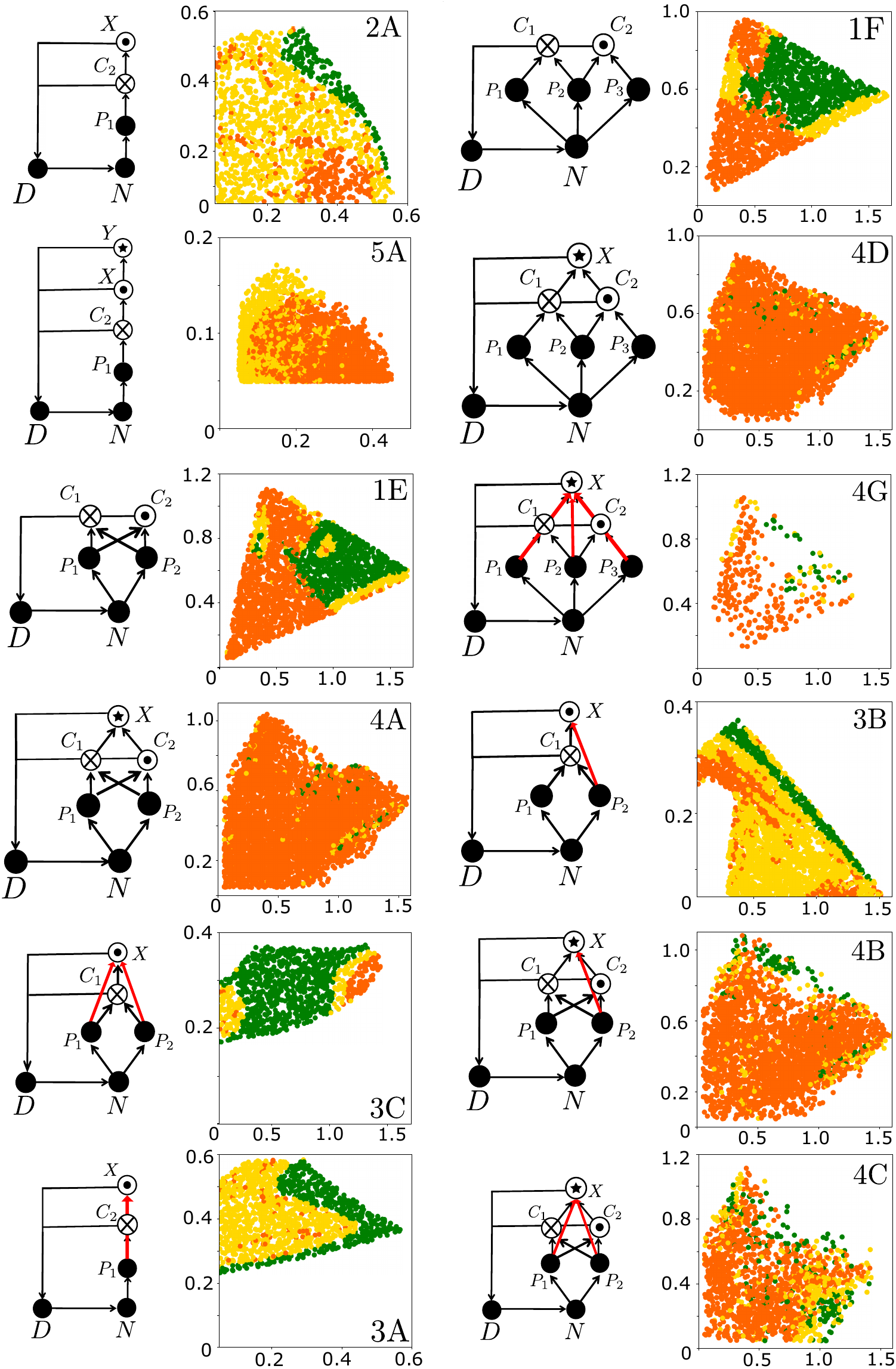}
    \caption{Dynamic behavior and graphical representation of key trophic network configurations. The $x-$ and $y-$ axes in the plots correspond respectively to the perturbations in the mortality rates of the ($\times$) and the ($\bullet$) state variables shown in the graphical representation. In the cases of three-parameter perturbations (5A, 4D, 4G, 4A, 4B, 4C) the third dimension (corresponding to the mortality rate of the ($\star$) state variable) is projected onto the $x-y$ plane. $Green \ points$ indicate steady states, $yellow \ points$ indicate periodic behavior, and $orange \ points$ indicate chaos. The $white \ area$ represents the extinction zone. $Red \ arrows$ in the graphic representation highlight omnivorous links.}
    \label{Results_comparazione}
\end
{figure}

\noindent As observed in the panels, increasing the number of consumers at the same trophic level destabilizes the system (Figure \ref{Summary Results}A). Statistical analyses show a clear increase in the median of the chaos percentage across consumer classes (1 consumer: 14.5 $\%$ ; 2 consumers 71 $\%$). On the other hand, omnivorous links tend to stabilize the network (Figure \ref{Summary Results}B). While clear patterns emerge at the level of individual groups of configurations, the statistical aggregate analysis shows no overall trend, with high variability, indicating that the degree of chaos, in this case, strongly depends on the specific configuration. In addition, increasing the number of trophic levels reduces the percentage of steady-state dynamics (Figure \ref{Summary Results}C), thereby amplifying chaotic behavior. This is also supported by statistical analysis, with the median steady-state percentage decreasing across trophic level classes (2 levels: 44.5 $\%$ ; 3 levels: 7.8 $\%$; 4–5 levels: $1.7 \%$). These trends are consistent across all configurations.
Moreover, we observed that chaotic configurations are not negligible and are, in many cases, the predominant dynamics. 
Finally, in Table \ref{table:Results} we note that an increasing number of consumers expands the coexistence region, whereas a higher number of trophic levels reduces it. \\
In Figure \ref{Results_comparazione}, we illustrate in detail the coexistence regions of a subset of the previously described configurations, with $green \ points$ representing steady \ states, $yellow \ points$ indicating periodic behavior, and $orange \ points$ corresponding to chaotic dynamics. The $white \ area$ represents the extinction zone. The $x-$ and $y-$ axes correspond, respectively, to perturbations in the mortality rates of the ($\times$) and the ($\bullet$) state variables. In the cases of three-parameter perturbations, the third dimension corresponds to the mortality rate of the ($\star$) state variable and is projected onto the $x-y$ plane.
All configurations, including those with three and four dimensions (i.e., the number of species whose mortality rates are treated as variable parameters), are plotted onto a two-dimensional space for clarity. The nomenclature is consistent with the one already presented (Figure \ref{fig:network_configurations} in \ref{Appendix: Nomenclature}). \\
We show that configuration $3C$, with one consumer, has a smaller coexistence region and is more stable than configuration $4C$, which includes two consumers (corresponding to Figure \ref{Summary Results}A, $brown \ points$). Similarly, configuration $3B$, with one consumer, has a smaller area and is more stable than configuration $4B$, which has two consumers (corresponding to Figure \ref{Summary Results}A, $purple \ squares$). Furthermore, omnivorous links play a stabilizing role in the food web. Configuration $4G$, which includes three omnivorous links ($red \ arrow$), is more stable than configuration $4D$ (corresponding to Figure \ref{Summary Results}B, $orange \ diamonds$). The low number of points in configuration $4G$ is due to the extinction of at least one species, as omnivorous links destabilize \citeauthor{levin1970community}'s (\citeyear{levin1970community}) rule.  Configuration $3C$, with two omnivorous links, is more stable than $3B$, which has only one omnivorous link (corresponding to Figure \ref{Summary Results}B, $green \ triangles$), while configuration $3A$, with one omnivorous link, is more stable than configuration $2A$, (corresponding to Figure \ref{Summary Results}B $yellow \ points$).\\
We also show that the 5-element chain $5A$ is much more unstable than the 4-element chain $2A$, consistent with  Figure \ref{Summary Results}C ($yellow \ points$).
Additionally, the longer $4A$ configuration is more unstable than $1E$ (corresponding to $light \ blue \ triangle$ in Figure \ref{Summary Results}C), and the longer $4D$ configuration is more unstable than $1F$ (corresponding to $light \ violet \ triangles$ in Figure \ref{Summary Results}C). \\
Finally, the most significant result is that the configurations are generally highly unstable and display significant percentages of chaos, reflecting intrinsic instability within the food web, as evident from all panels of Figure \ref{Summary Results}, all areas of Figure \ref{Results_comparazione}, and the data in Table \ref{table:Results}. \\
\edef\specialpage{\the\numexpr\value{page}+1\relax}
\AddToHook{shipout/after}{\ifnum\value{page}=\specialpage\relax
  \thispagestyle{empty}
\fi}

\section{Discussion and Conclusion}
\label{sec 4: Discussion}
\noindent Our eight-species model, which incorporates the microbial loop and mass conservation, demonstrates that longer food chains destabilize the system, while omnivorous links promote stability. These findings are consistent with results from previous studies \citep{holyoak1998omnivory, fussmann2002food, gross2005long}. Interestingly, increasing number of consumers destabilizes the system, with many configurations displaying significant oscillations and chaos. 
Predation facilitates phytoplankton coexistence by reducing competitive pressure for a single limiting nutrient and preventing dominant competitors from monopolizing it \citep{menge1976species, mccauley1979zooplankton}. Our findings support the idea that aperiodic oscillations in species abundance are an ecological mechanism sustaining species coexistence \citep{Karolyi2000, Roy2020, Mallmin2024,Occhipinti2024, Delabays2025}, especially when predator-prey interactions have a stronger role in characterizing the dynamics than competition \citep{Delabays2025}. \\
Our results show higher levels of chaos and oscillations than the simpler model without microbial loop \citep{fussmann2002food}. This result suggests that increasing model complexity and realism in this model increases instability. We argue that the complexity-stability relationship is not straightforward; several factors —such as the type of interspecific feedbacks or the presence of a limiting nutrient— can also play a crucial role \citep{barabas2017self}. The presence of a single shared nutrient may amplify coupled predator–prey oscillations, increasing the number of interacting species and promoting more complex dynamics and chaotic behavior, consistent with the findings of \cite{beninca2009coupled}. By contrast, in the simplified model \citep{fussmann2002food}, each phytoplankton species grows independently and is regulated by a quadratic logistic damping term, thereby limiting the complexity of interactions.  \cite{jansen2003complexity} argue that the mean and variance of interaction coefficients determine whether stability increases or decreases with complexity in ecological communities. Diamond trophic structures (e.g., configuration 4G, 4A) have been shown to display specific destabilizing responses when combined with climate-change effects. In particular, increasing temperature may destabilize the system below a given threshold, while further warming can lead to the extinction of higher trophic-level species and a consequent simplification, and potential re-stabilization of the food web \cite{kaur_persistence_2020}. Moreover, \cite{neutel2014interaction} show how the patterning of interaction strengths, feedback loops, and their positive/negative signs \citep{levins1974discussion, jacquet2016no} dominates over general network-topological properties in controlling ecosystem stability. This, however, assumes that interaction loops can be decoupled from network topology, even though network topology inevitably affects feedback loops.
\begin{figure}
    \hspace{-7mm}
    \centering
    \includegraphics[width=1.05\linewidth]{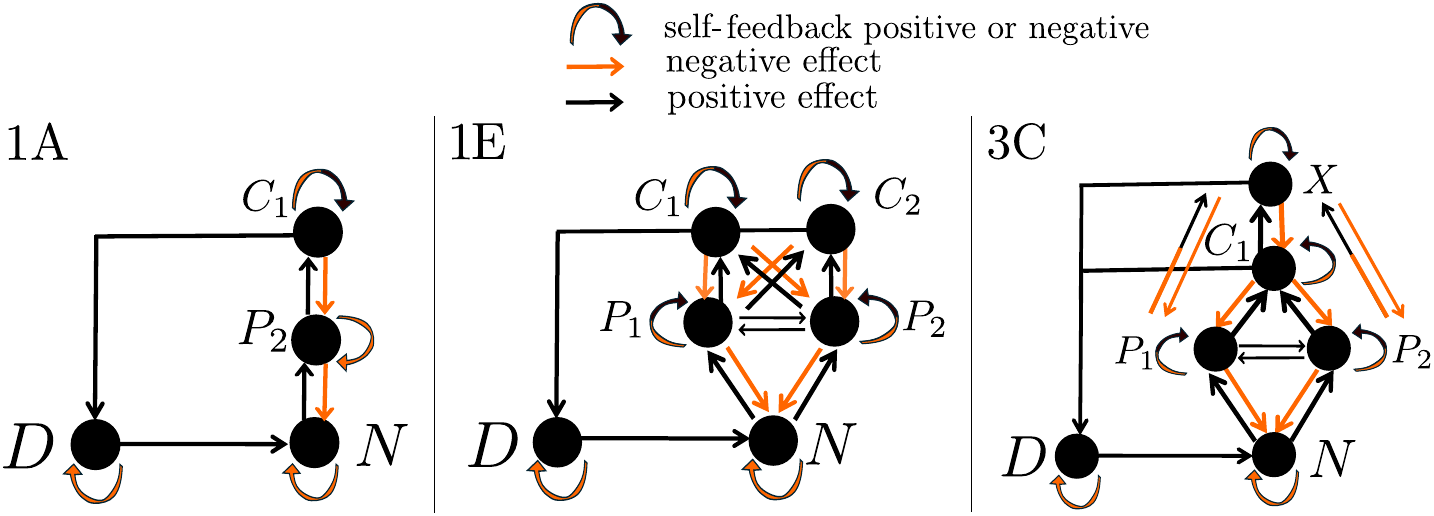}
    \caption{Feedback diagrams of configurations 1A ($left$), 1E ($center$), and 3C ($right$). $Black \ arrows$ indicate positive effects, while $orange \ arrows$  indicate negative effects. $Circular \  arrows$ represent self-effects. Some arrows are both black and orange because they may be either positive or negative depending on the variable values at a given equilibrium point. Conventions and representations of feedback according to Levins’ rules \citep{levins1974discussion}.}
    \label{fig:Diagramloops}
\end{figure}
\\
\noindent To investigate this, in Figure \ref{fig:Diagramloops} we present the feedback diagrams of selected configurations ($1A, 1E, 3C$), and we apply Levins’ rules \citep{levins1974discussion} to analyze the network stability in relation to feedback loops.  According to \cite{levins1974discussion}, a system of independent variables is stable when the interactions among feedback loops at each level—corresponding to the minors of the Jacobian— are strictly negative. This condition ensures that stabilizing, negative interactions dominate over destabilizing, positive ones. $Black \ arrows$ in Figure \ref{fig:Diagramloops} represent positive effects, while $orange \ arrows$ denote negative effects. Some arrows are both black and orange because they may be either positive or negative depending on the variable values at a given equilibrium point. Notably, in configuration 1E ($central$) and 3C ($right$), the presence of consumers promotes the formation of mutualistic effects among phytoplankton, facilitating their coexistence. In our system, total mass is conserved, and thus the sum of the system equations is identically zero. As a result, the rows of the Jacobian matrix are linearly dependent, which makes the Jacobian singular and leads to a zero determinant. Since Levins’ rules apply only to independent equations, we reformulate the system by expressing one variable as a function of the others, thereby reducing it to $n-1$ independent equations.
We then demonstrate that, at stable equilibria, all combinations of feedback loops $F_k$ are negative as further explained in \ref{Appendix: loopsanalysis}.  This analysis confirms that at steady state negative feedbacks dominate over positive ones across all levels, highlighting the importance of species interactions for trophic network stability. We also examine the second stability criterion, where for configuration 1A the condition $F_1 F_2 + F_3 > 0$ is satisfied, while for configurations 1E and 3C, the condition $ F_1^2 F_4 + F_1 F_5 - F_1 F_2 F_3 - F_3^2 > 0$ is proved. 
This suggests that feedback mechanisms across trophic levels are balanced, such that longer feedback loops do not dominate the stabilizing feedback generated by shorter loops.
Consistently, this analysis may also explain the greater instability of our network compared to the original one \citep{fussmann2002food}: with nutrient recycling we introduced positive interactions — detritus recycling and remineralization (represented by black arrows in Figure \ref{fig:Diagramloops}) — that counteract the effect of the stabilizing negative feedbacks.\\
Overall, our results show that the dynamical behavior of marine trophic networks is governed primarily by interactions and feedbacks among species — rather than complexity. By incorporating the microbial loop and enforcing mass conservation, the model reveals a wide range of dynamical regimes, including stable equilibria, periodic oscillations, and chaotic dynamics — which limits predictability. In particular, longer trophic chains and an increased number of consumers tend to destabilize the system, whereas omnivorous interactions promote stability. Compared with the baseline framework of \cite{fussmann2002food}, the inclusion of additional ecological processes leads to a  more unstable dynamical landscape, with frequent oscillatory and chaotic regimes.  In this context, analyzing interactions and feedback-loop dynamics may offer a valuable framework not only for diagnosing stability and chaos in ecological models, but also for prognostically understanding and anticipating the responses of marine ecosystems to climate change, with significant implications for ecological predictions and climate projections \citep{slingo2011uncertainty}.

\section*{Acknowledgements}
 \noindent This research was supported by the National Biodiversity Future Center NBFC project: National Recovery and Resilience Plan (NRRP), Mission 4 Component 2 Investment 1.4 – Call for tender No. 3138 of 16 December 2021, rectified by Decree no. 3175 of 18 December 2021 of Italian Ministry of University and Research funded by the European
Union – NextGenerationEU.
\clearpage
\centering

\appendix
\enlargethispage{2\baselineskip}
\section{\mbox{Model parameters}}
\label{Appendix: Model parameters}
\begin{table}[ht!]
\hspace{-1.25cm} 
\begin{tabular}{|c|l|c|}
\hline
\textbf{Parameter} & \textbf{Description} & \textbf{Value}  \\ \hline
$r_1$ & Growth rate of $P_1$ & 3 $  \text{day}^{-1}$ \\ \hline
$r_2$ & Growth rate of $P_2$ &  2 $\text{day}^{-1}$ \\ \hline
$r_3$ & Growth rate of $P_3$ &  1 $\text{day}^{-1}$ \\ \hline
$K_i$ & Half-saturation constant for $P_i$ with $i=1,2,3$ & 1 \\\hline
$a_{P_1}^{C_1}$ & Predation rate of $P_1$ by $C_1$ & 9 $\text{day}^{-1}$ \\ \hline
$a_{P_2}^{C_1}$ & Predation rate of $P_2$ by $C_1$ & 2 $\text{day}^{-1}$ \\ \hline
$a_{P_3}^{C_1}$ & Predation rate of $P_3$ by $C_1$ & 1 $\text{day}^{-1}$ \\ \hline
$a_{P_1}^{C_2}$ & Predation rate of $P_1$ by $C_2$ & 3 $\text{day}^{-1}$ \\ \hline
$a_{P_2}^{C_2}$ & Predation rate of $P_2$ by $C_2$ & 6 $\text{day}^{-1}$\\ \hline
$a_{P_3}^{C_2}$ & Predation rate of $P_3$ by $C_2$ & 3 $\text{day}^{-1}$\\ \hline
$b_{P_i}^{C_j}$ & Saturation constant of $P_i$ with $i=1,2,3$, by $C_j$ with $j=1,2$ & 5 \\ \hline
$a_{P_i}^{X}$ & Predation rate of $P_i$ with $i=1,2,3$, by $X$ & 0.25 $\text{day}^{-1}$ \\ \hline
$b_{P_i}^{X}$ & Saturation constant of $P_i$ with $i=1,2,3$ by $X$ & 0.5 \\ \hline
$b_{C_j}^{X}$ & Saturation constant of $C_1$ with $j=1,2$ by $X$ & 2.0 \\ \hline
$a_{C_j}^{X}$ & Predation rate of $C_j$ with $j=1,2$ by $X$ & 3 $\text{day}^{-1}$ \\ \hline
$b_{C_j}^{X}$ & Saturation constant of $C_j$ with $j=1,2$ by $X$ & 2.0\\ \hline
$a_{X}^{Y}$ & Predation rate of $X$ by $Y$ & 0.75 $\text{day}^{-1}$ \\ \hline
$b_{X}^{Y}$ & Saturation constant of $X$ by $Y$ & 0.5 \\ \hline
$a_{Y}^{Z}$ & Predation rate of $Y$ by $Z$ & 0.45 $\text{day}^{-1}$ \\ \hline
$b_{Y}^{Z}$ & Saturation constant of $Y$ by $Z$ & 0.3 \\ \hline
$d_{C_1}$ & Mortality term of $C_1$ & [0-2] $\text{day}^{-1}$\\ \hline
$d_{C_2}$ & Mortality term of $C_2$ & [0-2] $\text{day}^{-1}$\\ \hline
$d_X$ & Mortality term of $X$ & [0-2] $\text{day}^{-1}$
 \\ \hline
$d_Y$ & Mortality term of $Y$ & [0-2] $\text{day}^{-1}$ \\ \hline
$d_Z$ & Mortality term of $Z$ & [0-2] $\text{day}^{-1}$\\ \hline
$\tau$ & Remineralization time & 50 day \\ \hline
$P_{i0}$ & Initial condition of $P_i$ with $i=1,2,3$ & 0.5 \\ \hline
$C_{j0}$ & Initial condition of $C_j$ with $j=1,2$ & 0.5 \\ \hline
$X_0, Y_0, Z_0$ & Initial condition of $X$, $Y$, $Z$ & 0.5 \\ \hline
$D_0$ & Initial condition of $D$ & 2.0  \\ \hline
$N_0$ & Initial condition of $N$ & 2.0  \\ \hline
\end{tabular}
\caption{Model parameters.}
\label{tab:parameters}
\end{table}

\newpage
\section{Nomenclature of trophic network configurations}
\label{Appendix: Nomenclature}

\begin{figure}[ht!]
    \includegraphics[angle=90, width=0.99\linewidth]{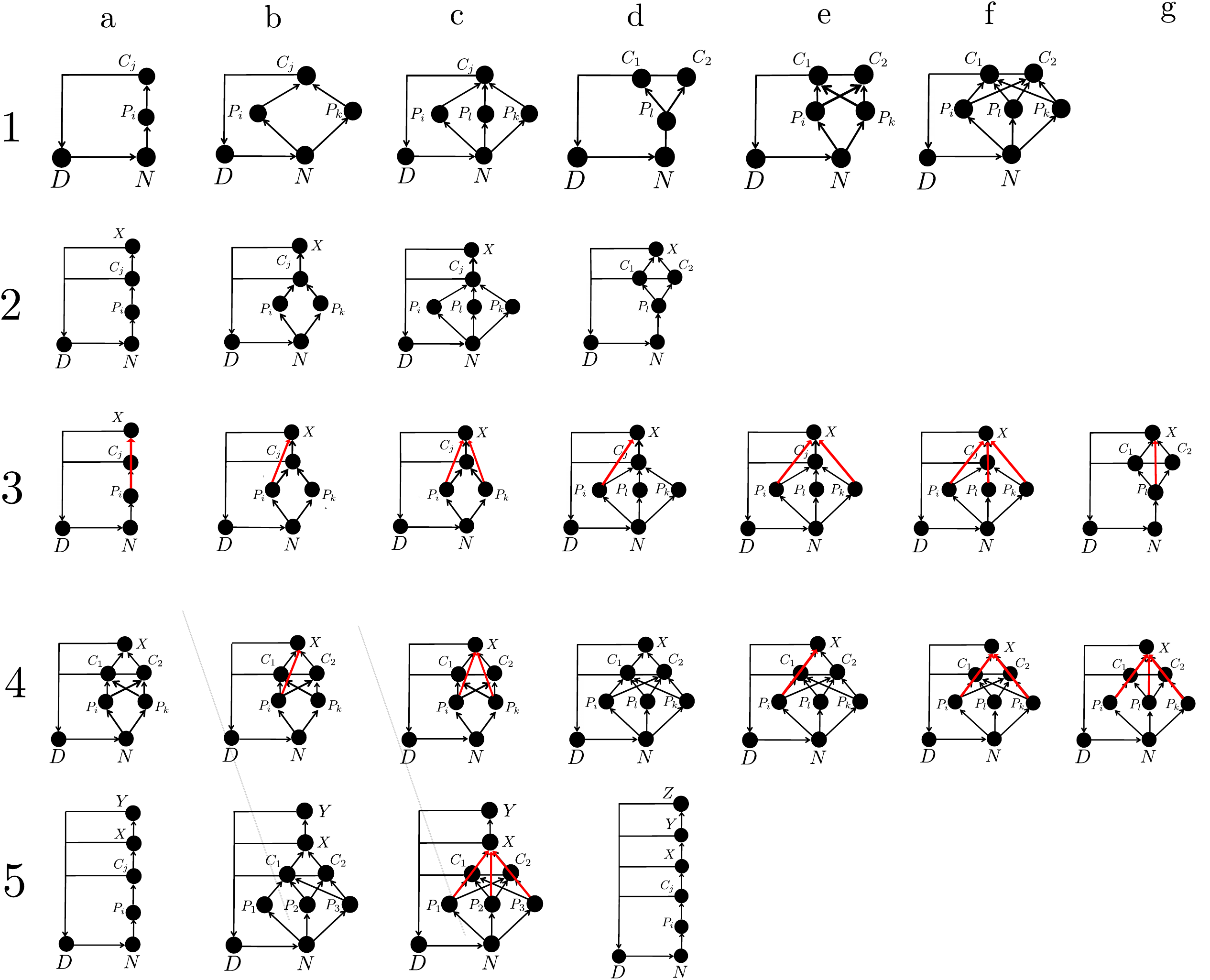}
    \caption{Nomenclature of trophic network configurations.}
    \label{fig:network_configurations}
\end{figure}

\section{Methodology}
\label{Appendix: Us vs Fussmann}
\begin{figure}[H]
    \centering
    \includegraphics[width=1.0\linewidth]{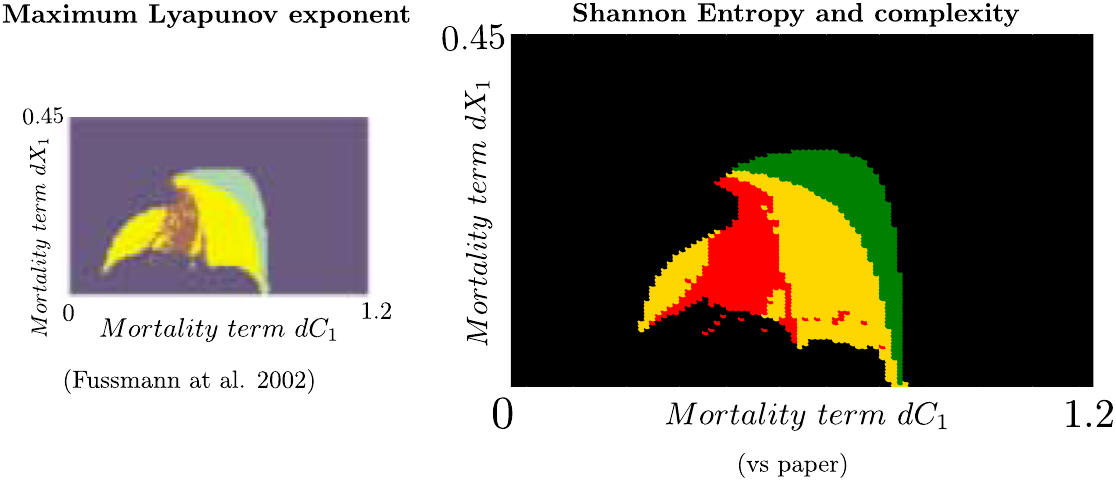}
    \caption{Comparison of methods for detecting chaos: On the left, \cite{fussmann2002food} used the traditional maximum Lyapunov exponent, while on the right, we applied the Shannon entropy and the Complexity Index. The methods yield consistent results.}
    \label{fig:comparisionFussmann}
\end{figure}
\label{app1}
\newpage

\section{Dynamic behavior of trophic network configurations}
\label{Appendix: Simulation results}

\setlength{\tabcolsep}{3pt}      
\renewcommand{\arraystretch}{0.9} 
\small                             

\setlength{\LTleft}{-1.5cm}   
\setlength{\LTright}{0cm}   
\begin{longtable}
{|p{3.8cm}|p{5cm}|p{2.0cm}|p{2.7cm}|p{2.2cm}|}
\hline
\textbf{CONFIGURATION} & \textbf{SPECIES} & \textbf{STEADY STATE \%} & \textbf{PERIODIC \%} & \textbf{CHAOS \%} \\
\hline
\endfirsthead

\hline
\textbf{CONFIGURATION} & \textbf{SPECIES} & \textbf{STEADY STATE \%} & \textbf{PERIODIC \%} & \textbf{CHAOS \%} \\
\hline
\endhead

\hline
\multicolumn{4.5}{r}{\textit{Continued on next page…}} \\
\endfoot

\hline
\endlastfoot

 1F & $C_1$,$C_2$,$P_1$,$P_2$,$P_3$ & 40.26 & 14.21 & 45.53 \\ \hline
1E & $P_1$,$P_3$,$C_1$,$C_2$ & 33.43 & 16.72 & 49.85 \\ \hline
1E & $P_1$,$P_2$,$C_1$,$C_2$ & 31.4  & 12.06 & 56.54 \\ \hline
1E & $P_2$,$P_3$,$C_1$,$C_2$ & 26.67 & 66.67 & 6.67 \\ \hline
1D & $P_2$,$C_1$,$C_2$ & \multicolumn{3}{|c|}{non-coexistence} \\ \hline
1D & $P_1$,$C_1$,$C_2$ & \multicolumn{3}{|c|}{non-coexistence} \\ \hline
1D & $P_3$,$C_1$,$C_2$ & \multicolumn{3}{|c|}{non-coexistence} \\ \hline
1C & $C_1$,$P_1$,$P_2$,$P_3$ & \multicolumn{3}{|c|}{non-coexistence} \\ \hline
1C & $C_2$,$P_1$,$P_2$,$P_3$ & \multicolumn{3}{|c|}{non-coexistence} \\ \hline
1B & $P_2$,$P_3$,$C_1$ & 63.16 & 36.84 & 0 \\ \hline
1B & $P_1$,$P_3$,$C_1$ & \multicolumn{3}{|c|}{non-coexistence} \\ \hline
1B & $P_1$,$P_2$,$C_1$ & \multicolumn{3}{|c|}{non-coexistence} \\ \hline
1B & $P_1$,$P_3$,$C_2$ & \multicolumn{3}{|c|}{non-coexistence} \\ \hline
1B & $P_1$,$P_2$,$C_2$ & \multicolumn{3}{|c|}{non-coexistence} \\ \hline
1B & $P_2$,$P_3$,$C_2$ & 51.67 & 45 & 3.33 \\ \hline
1A & $P_2$,$C_1$  & 50.88 & 49.12	& 0 \\ \hline
1A & $P_1$,$C_1$ & 44.54 & 53.64 & 1.82 \\ \hline
1A & $P_3$,$C_1$ & 50 & 50 & 0 \\ \hline
1A & $P_2$,$C_2$ & 39.73	& 60.27 &	0 \\ \hline
1A & $P_1$,$C_2$ & 57.14	& 42.86 & 0 \\ \hline
1A & $P_3$,$C_2$ & 31.43 & 65.71 & 2.86 \\ \hline
2A & $P_2$,$C_1$,$X$ & 7.2 & 57.04 & 35.77 \\ \hline
2A & $P_1$,$C_1$,$X$ & 15.66 & 76.55 & 7.79 \\ \hline
2A & $P_3$,$C_1$,$X$ & 2.78 & 29.42 & 67.8 \\ \hline
2A & $P_2$,$C_2$,$X$ & 10.82 & 81.38 & 7.8 \\ \hline
2A & $P_1$,$C_2$,$X$ & 11.68 & 67.32 & 21 \\ \hline
2A & $P_3$,$C_2$,$X$ & 5.2 & 74.47 & 20.33 \\ \hline
2B & $P_2$,$P_3$,$C_2$,$X$ & 16.29 & 74.67 & 9.04 \\ \hline
2B & $P_1$,$P_3$,$C_2$,$X$ & \multicolumn{3}{|c|}{non-coexistence} \\ \hline
2B & $P_1$,$P_2$,$C_2$,$X$ & \multicolumn{3}{|c|}{non-coexistence} \\ \hline
2B & $P_2$,$P_3$,$C_1$,$X$ & 8.31 & 46.88 & 44.81 \\ \hline
2B & $P_1$,$P_3$,$C_1$,$X$ & \multicolumn{3}{|c|}{non-coexistence} \\ \hline
2B & $P_1$,$P_2$,$C_1$,$X$ & \multicolumn{3}{|c|}{non-coexistence} \\ \hline
2C & $P_1$,$P_2$,$P_3$,$C_2$,$X$ & \multicolumn{3}{|c|}{non-coexistence} \\ \hline
2C & $P_1$,$P_2$,$P_3$,$C_1$,$X$ & \multicolumn{3}{|c|}{non-coexistence} \\ \hline
2D & $P_1$,$C_1$,$C_2$,$X$ & 33.76 & 29.64 & 36.6 \\ \hline
2D & $P_2$,$C_1$,$C_2$,$X$ & 23.91 & 18.84 & 57.25 \\ \hline
2D & $P_3$,$C_1$,$C_2$,$X$ & 5.69 & 8.13 & 86.18 \\ \hline
3A & $P_2$,$C_1$,$X$ & 28.03 & 67.35 & 4.62 \\ \hline
3A & $P_1$,$C_1$,$X$ & 66.08 & 32.51 & 1.41 \\ \hline
3A & $P_3$,$C_1$,$X$ & 17.44 & 74.38 & 8.18 \\ \hline
3A & $P_2$,$C_2$,$X$ & 59.29 & 36.34 & 4.37 \\ \hline
3A & $P_1$,$C_2$,$X$ & 34.41 & 61.47 & 4.12 \\ \hline
3A & $P_3$,$C_2$,$X$ & 60.99 & 35.87 & 3.14 \\ \hline
3B & $P_1$,$P_2$,$C_1$,$X$-omnivorous link with $P_1$ & \multicolumn{3}{|c|}{non-coexistence} \\ \hline
3B & $P_1$,$P_2$,$C_1$,$X$-omnivorous link with $P_2$ & 11.13 & 52.74 & 36.13 \\ \hline
3B & $P_1$,$P_3$,$C_1$,$X$-omnivorous link with $P_1$ & \multicolumn{3}{|c|}{non-coexistence} \\ \hline
3B & $P_1$,$P_3$,$C_1$,$X$-omnivorous link with $P_3$ & 6.34 & 57.22 & 36.44 \\ \hline
3B & $P_2$,$P_3$,$C_1$,$X$-omnivorous link with $P_2$ & 3.58 & 41.18 & 55.24 \\ \hline
3B & $P_2$,$P_3$,$C_1$,$X$-omnivorous link with $P_3$ & 8.39 & 63.87 & 27.74 \\ \hline
3B & $P_1$,$P_2$,$C_2$,$X$-omnivorous link with $P_1$ & 15.18 & 70.24 & 14.58 \\ \hline
3B & $P_1$,$P_2$,$C_2$,$X$-omnivorous link with $P_2$ & \multicolumn{3}{|c|}{non-coexistence} \\ \hline
3B & $P_1$,$P_3$,$C_2$,$X$-omnivorous link with $P_1$ & 26.67 & 53.53 & 19.8 \\ \hline
3B & $P_1$,$P_3$,$C_2$,$X$-omnivorous link with $P_3$ & \multicolumn{3}{|c|}{non-coexistence} \\ \hline
3B & $P_2$,$P_3$,$C_2$,$X$-omnivorous link with $P_2$ & 8.53 & 80.57 & 10.9 \\ \hline
3B & $P_2$,$P_3$,$C_2$,$X$-omnivorous link with $P_3$ & 25.58 & 65.12 & 9.3 \\ \hline
3C & $P_2$,$P_3$,$C_2$,$X$ & 60 & 33.33 & 6.67 \\ \hline
3C & $P_1$,$P_3$,$C_2$,$X$ & \multicolumn{3}{|c|}{non-coexistence} \\ \hline
3C & $P_1$,$P_2$,$C_2$,$X$ & \multicolumn{3}{|c|}{non-coexistence} \\ \hline
3C & $P_1$,$P_3$,$C_1$,$X$ & 57.2 & 35.52 & 7.27 \\ \hline
3C & $P_2$,$P_3$,$C_1$,$X$ & 14.42 & 73.08 & 12.5 \\ \hline
3C & $P_1$,$P_2$,$C_1$,$X$ & 74.73 & 15.86 & 9.41 \\ \hline
3D & $P_1$,$P_2$,$P_3$,$C_1$,$X$-omnivorous link with $P_1$ & \multicolumn{3}{|c|}{non-coexistence} \\ \hline
3D & $P_1$,$P_2$,$P_3$,$C_1$,$X$-omnivorous link with $P_2$ & \multicolumn{3}{|c|}{non-coexistence} \\ \hline
3D & $P_1$,$P_2$,$P_3$,$C_1$,$X$-omnivorous link with $P_3$ & \multicolumn{3}{|c|}{non-coexistence} \\ \hline
3D & $P_1$,$P_2$,$P_3$,$C_2$,$X$-omnivorous link with $P_1$ & 16.34 & 63.96 & 19.7 \\ \hline
3D & $P_1$,$P_2$,$P_3$,$C_2$,$X$-omnivorous link with $P_2$ & \multicolumn{3}{|c|}{non-coexistence} \\ \hline
3D & $P_1$,$P_2$,$P_3$,$C_2$,$X$-omnivorous link with $P_3$ & \multicolumn{3}{|c|}{non-coexistence} \\ \hline
3E & $P_1$,$P_2$,$P_3$,$C_1$,$X$-omnivorous links with $P_1$,$P_3$ & \multicolumn{3}{|c|}{non-coexistence} \\ \hline
3E & $P_1$,$P_2$,$P_3$,$C_1$,$X$-omnivorous links with $P_2$,$P_3$ & 60 & 38.18 & 1.82 \\ \hline
3E & $P_1$,$P_2$,$P_3$,$C_1$,$X$-omnivorous links with $P_1$,$P_2$ & \multicolumn{3}{|c|}{non-coexistence} \\ \hline
3E & $P_1$,$P_2$,$P_3$,$C_2$,$X$-omnivorous link with $P_1$,$P_3$ & 0 & 66.67 & 33.33 \\ \hline
3E & $P_1$,$P_2$,$P_3$,$C_2$,$X$-omnivorous links with $P_2$,$P_3$ & \multicolumn{3}{|c|}{non-coexistence} \\ \hline
3E & $P_1$,$P_2$,$P_3$,$C_2$,$X$-omnivorous links with $P_1$,$P_2$ & 11.11 & 88.89 & 0 \\ \hline
3F & $P_1$,$P_2$,$P_3$,$C_1$,$X$ & 62.5 & 37.5 & 0 \\ \hline
3F & $P_1$,$P_2$,$P_3$,$C_2$,$X$ & \multicolumn{3}{|c|}{non-coexistence} \\ \hline
3G & $P_1$,$C_1$,$C_2$,$X$-omnivorous link with $P_1$ & 50.85 & 32.2 & 16.95 \\ \hline
3G & $P_1$,$C_1$,$C_2$,$X$-omnivorous link with $P_2$ & 39.62 & 32.08 & 28.3 \\ \hline
3G & $P_1$,$C_1$,$C_2$,$X$-omnivorous link with $P_3$ & 50 & 10 & 40 \\ \hline
4A & $P_1$,$P_3$,$C_1$,$C_2$,$X$ & 4.09 & 9.55 & 86.36 \\ \hline
4A & $P_1$,$P_2$,$C_1$,$C_2$,$X$ & 1.45 & 3.38 & 95.17 \\ \hline
4A & $P_2$,$P_3$,$C_1$,$C_2$,$X$ & 29.41 & 15.81 & 54.78 \\ \hline
4B & $P_1$,$P_3$,$C_1$,$C_2$,$X$-omnivorous link with $P_1$ & 10.79 & 13.67 & 75.54 \\ \hline
4B & $P_1$,$P_3$,$C_1$,$C_2$,$X$-omnivorous link with $P_3$ & 6.92 & 23.9 & 69.18 \\ \hline
4B & $P_1$,$P_2$,$C_1$,$C_2$,$X$-omnivorous link with $P_1$ & 3.23 & 10.48 & 86.29 \\ \hline
4B & $P_1$,$P_2$,$C_1$,$C_2$,$X$-omnivorous link with $P_2$ & 5.82 & 5.56 & 88.62 \\ \hline
4B & $P_2$,$P_3$,$C_1$,$C_2$,$X$-omnivorous link with $P_2$ & 0 & 38.46 & 61.54 \\ \hline
4B & $P_2$,$P_3$,$C_1$,$C_2$,$X$-omnivorous link with $P_3$ & 20 & 40 & 40 \\ \hline
4C & $P_1$,$P_3$,$C_1$,$C_2$,$X$-omnivorous links with $P_1$,$P_3$ & 48.18 & 22.62 & 29.2 \\ \hline
4C & $P_1$,$P_2$,$C_1$,$C_2$,$X$-omnivorous links with $P_1$,$P_2$ & 14.47 & 22.13 & 63.4 \\ \hline
4C & $P_2$,$P_3$,$C_1$,$C_2$,$X$-omnivorous links with $P_2$,$P_3$ & 52.24 & 18.28 & 29.48 \\ \hline
4D & $P_1$,$P_2$,$P_3$,$C_1$,$C_2$,$X$ & 3.11 & 4 & 92.89 \\ \hline
4E & $P_1$,$P_2$,$P_3$,$C_1$,$C_2$,$X$-omnivorous link with $P_1$ & 5.88 & 10.3 & 83.82 \\ \hline
4E & $P_1$,$P_2$,$P_3$,$C_1$,$C_2$,$X$-omnivorous link with $P_2$ & 0 & 19.5 & 80.5 \\ \hline
4E & $P_1$,$P_2$,$P_3$,$C_1$,$C_2$,$X$-omnivorous link with $P_3$ & 8.2 & 5.74 & 86.06 \\ \hline
4F & $P_1$,$P_2$,$P_3$,$C_1$,$C_2$,$X$-omnivorous links with $P_1$,$P_3$ & 19.36 & 4.3 & 76.34 \\ \hline
4F & $P_1$,$P_2$,$P_3$,$C_1$,$C_2$,$X$-omnivorous links with $P_2$,$P_3$ & 12.5 & 9.87 & 77.63 \\ \hline
4F & $P_1$,$P_2$,$P_3$,$C_1$,$C_2$,$X$-omnivorous links with $P_1$,$P_2$ & 1.77 & 24.78 & 73.45 \\ \hline
4G & $P_1$,$P_2$,$P_3$,$C_1$,$C_2$,$X$-omnivorous links with all & 12.5 & 10.23 & 77.27 \\ \hline
5A & $P_1$,$C_1$,X,Y & 16.48 & 71.76 & 11.76 \\ \hline
5A & $P_2$,$C_1$,X,Y & 0 & 66.67 & 33.33 \\ \hline
5A & $P_3$,$C_1$,X,Y & \multicolumn{3}{|c|}{non-coexistence} \\ \hline
5A & $P_1$,$C_2$,X,Y & 0 & 39.13 & 60.87 \\ \hline
5A & $P_2$,$C_2$,X,Y & 0 & 68.09 & 31.91 \\ \hline
5A & $P_3$,$C_2$,X,Y & \multicolumn{3}{|c|}{non-coexistence} \\ \hline
5B & $P_1$,$P_2$,$P_3$,$C_1$,$C_2$,$X$,Y & 3.57 & 1.79 & 94.64 \\ \hline
5C & $P_1$,$P_2$,$P_3$,$C_1$,$C_2$,$X$,Y-omnivorous links with all & 3.83 & 6.51 & 89.66 \\ \hline
5D & $P_1$,$C_1$,X,Y,Z &3.48 &	20	& 76.52 \\ \hline
5D & $P_2$,$C_1$,X,Y,Z &0	& 9.52 & 	90.48 \\ \hline
5D & $P_3$,$C_1$,X,Y,Z &0	& 16.67 &	83.33 \\ \hline
5D & $P_1$,$C_2$,X,Y,Z & 5.41 & 8.11 & 86.49 \\ \hline
5D & $P_2$,$C_2$,X,Y,Z &3.33	& 33.33	& 63.33 \\ \hline
5D & $P_3$,$C_2$,X,Y,Z &0	& 12.5 	& 87.5 \\ \hline
\caption{Percentage distribution of steady-state (SS), periodic, and chaotic dynamics across all configurations of the trophic network.}
\label{table:Results}
\end{longtable}

\newpage
\section{\mbox{Feedback loops analysis}}
\label{Appendix: loopsanalysis}
Here we calculate the feedbacks $F_k$ of the 1A configuration according to Levins’ rules \citep{levins1974discussion}. The minors of the Jacobian — that is, the interactions among loops at all levels — must be negative. We calculate $F_k$ as: \\[0.25cm]
(feedback \ length\ 1, $self-damping \ loops$)\\[0.01cm]
$F_1= \gamma_{DD} + \gamma_{NN} + \gamma_{PP}+ \gamma_{CC}$ \\[0.25cm]
(feedback \ length \ 2): \\

$F_2= \gamma_{PN}\gamma_{NP} + \gamma_{CP}\gamma_{PC} - \gamma_{DD}\gamma_{NN}-\gamma_{DD}\gamma_{PP}- \gamma_{DD}\gamma_{CC}- \gamma_{NN}\gamma_{PP}- \gamma_{NN}\gamma_{CC}-
\gamma_{PP}\gamma_{CC} $ \\[0.25cm]

(feedback \ length \ 3): \\
$F_3= -(\gamma_{PN}\gamma_{NP})\gamma_{CC} -(\gamma_{PN}\gamma_{NP})\gamma_{DD} -(\gamma_{CP}\gamma_{PC})\gamma_{DD} -(\gamma_{CP}\gamma_{PC})\gamma_{NN} + \gamma_{DD}\gamma_{NN}\gamma_{PP} + \gamma_{DD}\gamma_{NN}\gamma_{CC} +
\gamma_{NN}\gamma_{PP}\gamma_{CC} +
\gamma_{PP}\gamma_{CC}\gamma_{DD}$  \\[0.25cm]
(feedback \ length \ 4): \\
$F_4=(\gamma_{ND}\gamma_{PN})\gamma_{CP}\gamma_{DC} + (\gamma_{PN}\gamma_{NP})\gamma_{DD}\gamma_{CC} + (\gamma_{PC}\gamma_{CP})\gamma_{NN}\gamma_{DD} -
\gamma_{PP}\gamma_{CC}\gamma_{DD}\gamma_{CC}$  \\[0.25cm]

Substituting into the equilibrium points, we obtain
\begin{equation}
F_1 < 0, \qquad
F_2 < 0, \qquad
F_3 < 0, \qquad
F_4 = 0 .
\end{equation}

The condition $F_4 = 0$ is due to mass conservation, implying that the variables are not independent. 
To address this issue, we express $D$ as a function of the other variables:
\begin{equation}
D = M - P_2 - C_1 - N ,
\end{equation}
where $M$ denotes the total mass.

In this way, we obtain a reduced $3 \times 3$ matrix, where variables are independent and verify that:
\begin{equation}
F_1 < 0, \qquad F_2 < 0, \qquad F_3 < 0
\end{equation}
demonstrating that negative feedbacks dominate positive ones at steady state. The analysis of all other configurations follows the same approach, is omitted and left to the reader.

\newpage
\bibliographystyle{plainnat}
\bibliography{thebibliography}

@article{fussmann2002food,
  title={Food web complexity and chaotic population dynamics},
  author={Fussmann, Gregor F and Heber, Gerd},
  journal={Ecology Letters},
  volume={5},
  number={3},
  pages={394--401},
  year={2002},
  publisher={Wiley Online Library}
}

@article{zunino2012distinguishing,
  title={Distinguishing chaotic and stochastic dynamics from time series by using a multiscale symbolic approach},
  author={Zunino, Luciano and Soriano, Miguel C and Rosso, Osvaldo A},
  journal={Physical Review E—Statistical, Nonlinear, and Soft Matter Physics},
  volume={86},
  number={4},
  pages={046210},
  year={2012},
  publisher={APS}
}

@article{rosso2007distinguishing,
  title={Distinguishing noise from chaos},
  author={Rosso, Osvaldo A and Larrondo, HA and Martin, Mar{\'\i}a Teresa and Plastino, A and Fuentes, Miguel A},
  journal={Physical review letters},
  volume={99},
  number={15},
  pages={154102},
  year={2007},
  publisher={APS}
}

@article{hemmings2004parameterizing,
  title={Parameterizing the microbial loop: an experiment in reducing model complexity},
  author={Hemmings, JCP and Srokosz, MA and Challenor, PG and Fasham, MJR},
  year={2004},
  publisher={Southampton Oceanography Centre}
}

@article{levin1970community,
  title={Community equilibria and stability, and an extension of the competitive exclusion principle},
  author={Levin, Simon A},
  journal={The American Naturalist},
  volume={104},
  number={939},
  pages={413--423},
  year={1970},
  publisher={University of Chicago Press}
}

@misc{bolding2020parsac,
  title={Parsac: parallel sensitivity analysis and calibration},
  author={Bolding, K and Bruggeman, J},
  year={2020}
}

@article{bruggeman2014general,
  title={A general framework for aquatic biogeochemical models},
  author={Bruggeman, Jorn and Bolding, Karsten},
  journal={Environmental modelling \& software},
  volume={61},
  pages={249--265},
  year={2014},
  publisher={Elsevier}
}

@article{crutchfield1986chaos,
  title={Chaos Scientific American 225 (6): 46-57},
  author={Crutchfield, J and Farmer, J and Packard, N and Shaw, R},
  journal={Crutchfeld646225Scientific American},
  year={1986}
}

@article{rogers2022chaos,
  title={Chaos is not rare in natural ecosystems},
  author={Rogers, Tanya L and Johnson, Bethany J and Munch, Stephan B},
  journal={Nature ecology \& evolution},
  volume={6},
  number={8},
  pages={1105--1111},
  year={2022},
  publisher={Nature Publishing Group UK London}
}

@article{azam2007microbial,
  title={Microbial structuring of marine ecosystems},
  author={Azam, Farooq and Malfatti, Francesca},
  journal={Nature Reviews Microbiology},
  volume={5},
  number={10},
  pages={782--791},
  year={2007},
  publisher={Nature Publishing Group UK London}
}

@article{altman2018curse,
  title={The curse (s) of dimensionality},
  author={Altman, Naomi and Krzywinski, Martin},
  journal={Nat Methods},
  volume={15},
  number={6},
  pages={399--400},
  year={2018}
}

@article{barbier2019pyramids,
  title={Pyramids and cascades: a synthesis of food chain functioning and stability},
  author={Barbier, Matthieu and Loreau, Michel},
  journal={Ecology letters},
  volume={22},
  number={2},
  pages={405--419},
  year={2019},
  publisher={Wiley Online Library}
}

@article{pessa2021ordpy,
  title={ordpy: A Python package for data analysis with permutation entropy and ordinal network methods},
  author={Pessa, Arthur AB and Ribeiro, Haroldo V},
  journal={Chaos: An Interdisciplinary Journal of Nonlinear Science},
  volume={31},
  number={6},
  year={2021},
  publisher={AIP Publishing}
}

@article{occhipinti2023marine,
  title={Marine ecosystem models of realistic complexity rarely exhibits significant endogenous non-stationary dynamics},
  author={Occhipinti, Guido and Solidoro, Cosimo and Grimaudo, Roberto and Valenti, Davide and Lazzari, Paolo},
  journal={Chaos, Solitons \& Fractals},
  volume={175},
  pages={113961},
  year={2023},
  publisher={Elsevier}
}

@article{smale1976differential,
  title={On the differential equations of species in competition},
  author={Smale, Steve},
  journal={Journal of Mathematical Biology},
  volume={3},
  number={1},
  pages={5--7},
  year={1976},
  publisher={Springer Science and Business Media Deutschland GmbH}
}

@article{barabas2017self,
  title={Self-regulation and the stability of large ecological networks},
  author={Barab{\'a}s, Gy{\"o}rgy and Michalska-Smith, Matthew J and Allesina, Stefano},
  journal={Nature ecology \& evolution},
  volume={1},
  number={12},
  pages={1870--1875},
  year={2017},
  publisher={Nature Publishing Group UK London}
}

@article{berryman1989ecological,
  title={Are ecological systems chaotic—and if not, why not?},
  author={Berryman, AA and Millstein, JA},
  journal={Trends in Ecology \& Evolution},
  volume={4},
  number={1},
  pages={26--28},
  year={1989},
  publisher={Elsevier}
}

@article{roy2019numerical,
  title={Numerical implementation of dynamical mean field theory for disordered systems: Application to the Lotka--Volterra model of ecosystems},
  author={Roy, Felix and Biroli, Giulio and Bunin, Guy and Cammarota, Chiara},
  journal={Journal of Physics A: Mathematical and Theoretical},
  volume={52},
  number={48},
  pages={484001},
  year={2019},
  publisher={IOP Publishing}
}

@article{gross2005long,
  title={Long food chains are in general chaotic},
  author={Gross, Thilo and Ebenh{\"o}h, Wolfgang and Feudel, Ulrike},
  journal={Oikos},
  volume={109},
  number={1},
  pages={135--144},
  year={2005},
  publisher={Wiley Online Library}
}

@article{holyoak1998omnivory,
  title={Omnivory and the stability of simple food webs},
  author={Holyoak, Marcel and Sachdev, Sambhav},
  journal={Oecologia},
  volume={117},
  number={3},
  pages={413--419},
  year={1998},
  publisher={Springer}
}

@article{beninca2009coupled,
  title={Coupled predator--prey oscillations in a chaotic food web},
  author={Beninc{\`a}, Elisa and J{\"o}hnk, Klaus D and Heerkloss, Reinhard and Huisman, Jef},
  journal={Ecology letters},
  volume={12},
  number={12},
  pages={1367--1378},
  year={2009},
  publisher={Wiley Online Library}
}

@article{mccauley1979zooplankton,
  title={Zooplankton grazing and phytoplankton species richness: Field tests of the predation hypothesis 1},
  author={McCauley, Edward and Briand, Fr{\'e}d{\'e}ric},
  journal={Limnology and Oceanography},
  volume={24},
  number={2},
  pages={243--252},
  year={1979},
  publisher={Wiley Online Library}
}

@article{menge1976species,
  title={Species diversity gradients: synthesis of the roles of predation, competition, and temporal heterogeneity},
  author={Menge, Bruce A and Sutherland, John P},
  journal={The American Naturalist},
  volume={110},
  number={973},
  pages={351--369},
  year={1976},
  publisher={University of Chicago Press}
}

@article{jansen2003complexity,
  title={Complexity and stability revisited},
  author={Jansen, Vincent AA and Kokkoris, Giorgos D},
  journal={Ecology letters},
  volume={6},
  number={6},
  pages={498--502},
  year={2003},
  publisher={Wiley Online Library}
}

@article{jacquet2016no,
  title={No complexity--stability relationship in empirical ecosystems},
  author={Jacquet, Claire and Moritz, Charlotte and Morissette, Lyne and Legagneux, Pierre and Massol, Fran{\c{c}}ois and Archambault, Philippe and Gravel, Dominique},
  journal={Nature communications},
  volume={7},
  number={1},
  pages={12573},
  year={2016},
  publisher={Nature Publishing Group UK London}
}

@article{neutel2014interaction,
  title={Interaction strengths in balanced carbon cycles and the absence of a relation between ecosystem complexity and stability},
  author={Neutel, Anje-Margriet and Thorne, Michael AS},
  journal={Ecology letters},
  volume={17},
  number={6},
  pages={651--661},
  year={2014},
  publisher={Wiley Online Library}
}

@article{slingo2011uncertainty,
  title={Uncertainty in weather and climate prediction},
  author={Slingo, Julia and Palmer, Tim},
  journal={Philosophical Transactions of the Royal Society A: Mathematical, Physical and Engineering Sciences},
  volume={369},
  number={1956},
  pages={4751--4767},
  year={2011},
  publisher={The Royal Society Publishing}
}

@article{Delabays2025, title={Route to chaos in multi-species ecosystems}, volume={35}, ISSN={1054-1500, 1089-7682}, DOI={10.1063/5.0291485},  number={9}, journal={Chaos: An Interdisciplinary Journal of Nonlinear Science}, author={Delabays, Robin and Jacquod, Philippe}, year={2025}, pages={091109}, language={en} }

@article{Karolyi2000, title={Chaotic flow: The physics of species coexistence}, volume={97}, ISSN={0027-8424, 1091-6490}, DOI={10.1073/pnas.240242797},  number={25}, journal={Proceedings of the National Academy of Sciences}, author={Károlyi, György and Péntek, Áron and Scheuring, István and Tél, Tamás and Toroczkai, Zoltán}, year={2000},  pages={13661–13665}, language={en} }

@article{Mallmin2024, title={Chaotic turnover of rare and abundant species in a strongly interacting model community}, volume={121}, ISSN={0027-8424, 1091-6490}, DOI={10.1073/pnas.2312822121},  number={11}, journal={Proceedings of the National Academy of Sciences}, author={Mallmin, Emil and Traulsen, Arne and De Monte, Silvia}, year={2024}, pages={e2312822121}, language={en} }

@article{Occhipinti2024, title={Stochastic effects on plankton dynamics: Insights from a realistic 0-dimensional marine biogeochemical model}, volume={83}, ISSN={15749541}, DOI={10.1016/j.ecoinf.2024.102778}, journal={Ecological Informatics}, author={Occhipinti, Guido and Piani, Stefano and Lazzari, Paolo}, year={2024}, pages={102778}, language={en} }

@article{Roy2020, title={Complex interactions can create persistent fluctuations in high-diversity ecosystems}, volume={16}, ISSN={1553-7358}, DOI={10.1371/journal.pcbi.1007827}, number={5}, journal={PLOS Computational Biology}, author={Roy, Felix and Barbier, Matthieu and Biroli, Giulio and Bunin, Guy}, editor={Grilli, Jacopo}, year={2020},  pages={e1007827}, language={en} }

@article{levins1974discussion,
  title={Discussion paper: the qualitative analysis of partially specified systems},
  author={Levins, Richard},
  journal={Annals of the New York Academy of Sciences},
  volume={231},
  number={1},
  pages={123--138},
  year={1974},
  publisher={Wiley Online Library}
}

@article{cohen1993body,
  title={Body sizes of animal predators and animal prey in food webs},
  author={Cohen, Joel E and Pimm, Stuart L and Yodzis, Peter and Salda{\~n}a, Joan},
  journal={Journal of animal ecology},
  pages={67--78},
  year={1993},
  publisher={JSTOR}
}

@article{kaur_persistence_2020,
	title = {Persistence and stability of interacting species in response to climate warming: the role of trophic structure},
	volume = {13},
	issn = {1874-1738, 1874-1746},
	shorttitle = {Persistence and stability of interacting species in response to climate warming},
	url = {https://link.springer.com/10.1007/s12080-020-00456-9},
	doi = {10.1007/s12080-020-00456-9},
	language = {en},
	number = {3},
	urldate = {2026-05-26},
	journal = {Theoretical Ecology},
	author = {Kaur, Taranjot and Dutta, Partha Sharathi},
	month = sep,
	year = {2020},
	pages = {333--348},
}

\end{document}